# Equilibrium structures of water molecules confined within a multiply connected carbon nanotube: a molecular dynamics study

**Taehoon Kim** [a], **Gwan Woo Kim** [b], **Hyunah Jeong** [a], **Gunn Kim** *[b] **and Soonmin Jang** *[a]

[a]*Department of Chemistry, Sejong University, Seoul 05006, Korea.*

[b]*Department of Physics and Astronomy & Institute of Fundamental Physics, Sejong University, Seoul 05006, Korea*

E-mail: gunnkim@sejong.ac.kr; sjang@sejong.ac.kr

**Abstract**

Water confinement inside a carbon nanotube (CNT) has been one of the most exciting subjects of both experimental and theoretical interest. Most of the previous studies, however, considered CNT structures with simple cylindrical shapes. In this paper, we report a classical molecular dynamics study of the equilibrium structural arrangement of water molecules confined in a multiply connected carbon nanotube (MCCNT) containing two Y-junctions. We investigate the structural arrangement of the water molecules in the MCCNT in terms of the density of water molecules and the average number of hydrogen bonds per water molecule. Our results show that the structural rearrangement of the $H_2O$ molecules takes place several angstroms ahead of the Y-junction, rather than only at the CNT junction itself. This phenomenon arises because it is difficult to match the boundary condition for hydrogen bonding in the region where two different hydrogen-bonded structures are interconnected with each other.

**Introduction**

The highly favourable filling of a carbon nanotube (CNT) with water molecules due to a capillary effect is a well-known phenomenon and has been the subject of numerous studies both experimentally and theoretically.[1,2] Recently, molecular dynamics (MD) simulations have revealed that the driving force and the detailed structural arrangement of the water molecules participating in this favourable CNT wetting depends on the diameter size of the tube.[3] The difference of the water nanostructure, the structural network of water molecules to be more specific, within the CNT is not new because the environment of water molecules changes from that of nano-sized water confinement to that of bulk water as the tube diameter increases. Some researchers showed structural and dynamic features of water confined within the CNT.[4] For example, the hydrogen bond of water inside a CNT is weak compared to bulk water[5] as predicted from an MD study.[6,7] A detailed understanding of specific molecules confined inside a well-characterized nanosystem provides valuable information such as energy, structure, and transport properties of the corresponding materials.[8] In this respect, the water nanostructures inside the CNT may serve as templates for the effective design of a nanochannel for various practical purposes such as water purification[9] and drug delivery[10,11] to name a few.

Understanding how the physical properties of the fluid can be modified[12] by introducing non-carbon atoms or functional groups to the host structure has crucial importance in controlling the static and dynamic transport phenomena of fluid such as flow rate at the nanoscale.[13,14] Recent studies have shown that the local cross-sectional area of the microchannel can change for controlling the rate of capillary flow.[15–20] Therefore, the use of structurally inhomogeneous hosts, such as a CNT with varying local cross-sectional area, might be a possible strategy to control the nanofluid as a channel. The water molecules in complex nanostructures that connect nanotubes of different diameters could show very intriguing structural features. The behaviour of water molecules in topologically complicated CNTs has not yet been reported, whereas reports of wetting studies of non-carbon nanotubes, such as boron nitride nanotubes, have been published.[21,22]

In this paper, we report an MD study of the behaviour of water molecules inside a multiply connected carbon nanotubes (MCCNT).[23,24] MCCNTs could be produced by CNTs of different sizes clinging together under the influence of electronic beams to form Y-junctions,[25–28] in which a large CNT splits into two smaller CNT arms that eventually merge into one. Our study elucidates the structural arrangement of the water molecules inside these topologically complicated CNTs in detail, especially in the vicinity of the junction at which a CNT with a larger diameter is connected to two CNTs with a smaller diameter.

**Computational methods**

As shown in Fig. 1, we consider an MCCNT with a topologically complicated symmetry, of which the (14, 14) CNT branches into two (7, 7) CNTs, which are then fused into the (14, 14) CNT. Because of the bifurcated shape, each Y-junction in the MCCNT contains six heptagons, and the entire system is mirror-symmetric about two planes, M1 and M2. Here, M1 is a plane that bisects the MCCNT along the axis direction of the tube, and M2 is a plane that is perpendicular to M1 and bisects the MCCNT. The side view shows the MCCNT along the M2 plane, and the cross-sectional view shows the MCCNT along the M1 plane. If looking at the cross-section, one can see that the entrance of the MCCNT looks elliptical. If the (14, 14) CNT part were much longer, the part would have a circular cross-section. Near the junction, however, it should be deformed to connect naturally to the (7, 7) CNTs with a smaller diameter. The overall length of the CNT is ~8 nm. The distance from the left end of the CNT to the position of the left Y-junction is 24.1 Å. Because the Y-junction connects two nanotubes with different sizes, the shape of the entire CNT becomes slightly flattened as the position changes from an open end of the (14, 14) CNT to its interior as is clear from the side view of the MCCNT.

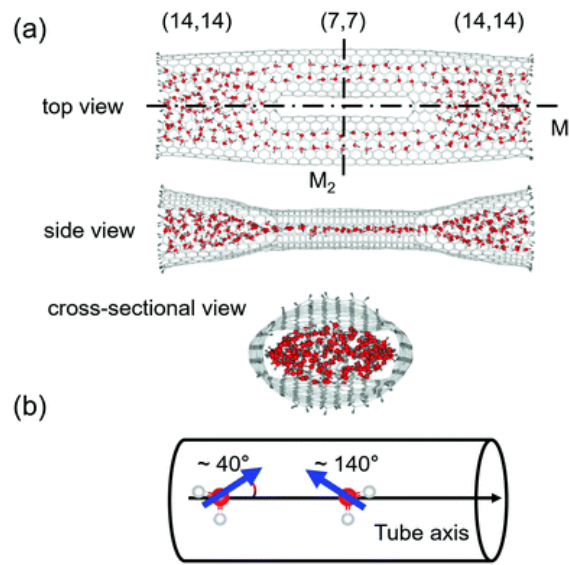

**Fig. 1** (a) A model structure of the MCCNT that contains the water molecules for the TIP3P model (a snapshot at 20 ns). (b) Schematic illustration of the orientation of the dipole moment of an $H_2O$ molecule.

We initially put the MCCNT in a cubic box with a minimum distance of 15.0 Å from the edge of the CNT to the wall of the box and filled the box with water molecules. After solvation and energy minimization, the entire system was subject to constant pressure and constant temperature (*NPT*) molecular dynamics simulation for 20 ns. In our work, the pressure was set to 1 bar at two different temperatures, *i.e.*, 280 K and 300 K, respectively, to observe the temperature-dependent behaviour. Usually, researchers investigate the properties of water at room temperature (300 K). However, considering the increased density of water as the temperature decreases, it seems very interesting to study what structural properties the water has at low temperatures. Thus, we also studied the properties of water molecules nanostructures confined to nanotubes at 280 K. In the 20 ns MD simulation, the initial 10 ns calculation shows a process of equilibration, so only the MD trajectories after 10 ns were used in this study. After initial solvation, the interior of the CNT is quickly filled with water molecules, usually within several tens of picoseconds. The simulation time step was 1 fs. We collected simulation trajectories every 20 steps and generated a total of 5 × $10^5$ frames for final analysis. The water molecules were allowed to enter or to exit the CNT freely, contacting with the water molecules in the bath.[2]

We used the modified Berendsen and isotropic Berendsen methods for temperature and pressure control with the control parameters 0.1 and 16.0 ps, respectively. The overall simulation was performed using the Gromacs ver. 5.1.2[29] with the CNT force field parameters taken from the OPLS-AA[30] and the TIP3P water model.[31] We also used the TIP4P water model[31] to investigate model dependence.[32] The charge of both C and H atoms in the CNT is set to be neutral in our model.[33] We used the leap-frog integrator for time integration. This scheme processes the Lennard-Jones parameters between a carbon atom in the CNT and an oxygen atom in water by using the Lorentz–Berthelot mixing rule. We also performed the same simulations with the perfect (14, 14) and (7, 7) CNTs with the tube length of 60 Å for comparison. The radii of the (7, 7) and (14, 14) tubes are 4.75 Å and 9.50 Å, respectively. All simulations were performed with periodic boundary conditions in the *x*-, *y*-, and *z*-directions with a short-range cutoff distance of 10 Å. The long-range electrostatic interaction was calculated by the particle mesh Ewald method (PME) with 1.6 Å grid spacing.

**Results and discussion**

Usually, the duration of a simulation of a typical CNT/water system to study its equilibrium properties is several nanoseconds.[1] Our estimation of the number of water molecules inside our CNT models as a function of the simulation time, as shown in Fig. 2, showed that they remain steady throughout the simulation with some fluctuations. Although this is not the sole criterion for equilibrium, we believe our sampling of a 10 ns trajectory after the initial 10 ns of simulation time to be sufficiently reasonable to study the equilibrium water configuration inside the CNTs. Fig. 3 shows a snapshot of the water molecules contained in the MCCNT at the end of the simulation. We emphasize that water molecules line up in two rows in the (7, 7) CNT region, whereas they form a cluster-like structure in the (14, 14) CNT region.

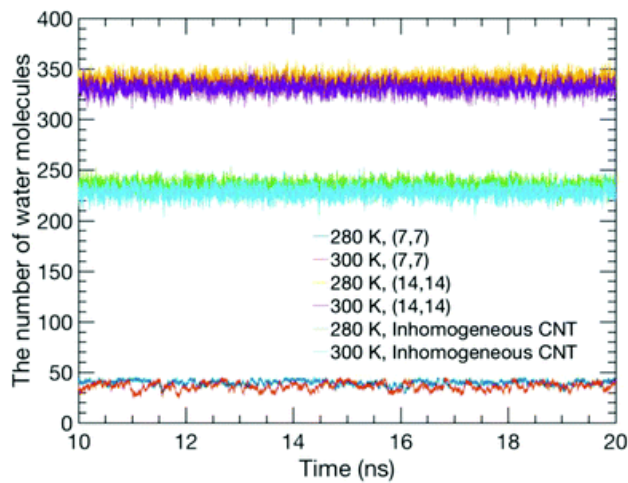

**Fig. 2** Number of water molecules inside the CNTs as a function of simulation time for the TIP3P model. The results from the TIP4P model are similar to those from the TIP3P.

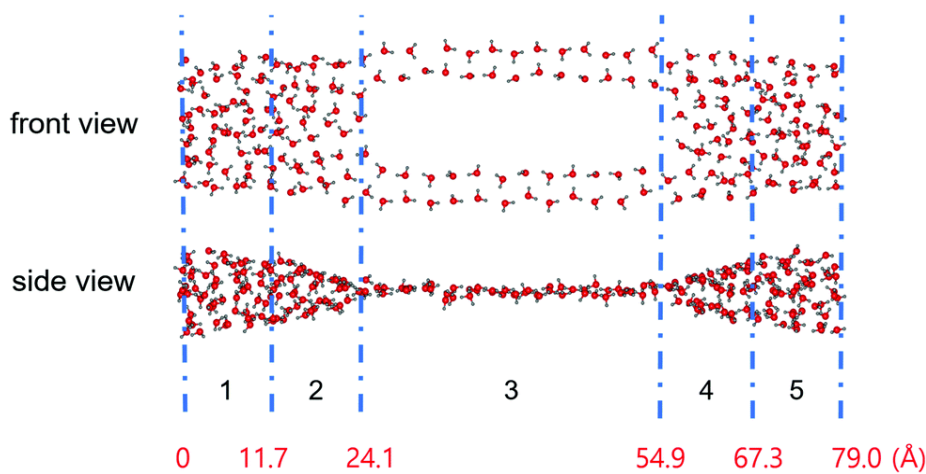

**Fig. 3** Snapshot of the result of simulating the behaviour of water molecules in the MCCNT at 280 K for 20 ns with TIP3P water. The MCCNT structure is omitted for clarity. The blue dashed line separates the regions, as indicated in Table 1. Numbers in red indicate the distance from the left end of the MCCNT.

Interestingly, unlike in the (14, 14) CNT, the electric dipole moment in water molecules tends to be orientated at an angle of either 40° or 140° relative to the direction of the nanotube axis in the (7, 7) CNT, as clearly shown in Fig. 1. The fact that one of the hydrogen atoms in a water molecule prefers to point to the hexagonal centre of graphene or CNT has already been reported by researchers who studied the structural properties of water molecules adsorbed inside a CNT or on a graphene sheet.[34–38] Because the water molecules can enter both open ends of the MCCNT, the angles of 40° or 140° are equivalent.

Fig. 3 clearly shows that water molecules confined inside the MCCNT have different characteristics in various regions. In Region 1, the water molecules form an elliptical column. In Region 2, which constitutes the Y-junction where the smaller (7, 7) CNTs connect to the larger (14, 14) CNT, the MCCNT is deformed.

In this region, therefore, the shape of the water nanostructure in the CNTs is also distorted. Roughly speaking, in Region 3, the double-stranded water nanostructure is on a plane. We calculated the numbers ($N_S$) of water molecules near the inner surface of the nanotubes and the total numbers ($N_V$) of water molecules in each region of the water nanostructure (Table 1). The results reveal a dimension-dependent phenomenon. Physical features of the water molecules confined inside the MCCNT can be characterized by the number of total water molecules divided by the number of water molecules near the inner surface of the CNT, *i.e.*, $N_S/N_V$ ratio. We define the water molecules as surface water molecules when the oxygen atoms of the water molecules are within 3.9 Å from the CNT surface. The detailed geometry of the carbon nanotube and many other complex factors affect the details of the water molecule network structure, resulting in dimensional dependence.

**Table 1** Number ($N_V$) of the total water molecules and number ($N_S$) of the water molecules at the surface of CNT and the ratio of $N_S/N_V$ depending on MCCNT regions for the TIP3P model

| Region | 1 | 2 | 3 | 4 | 5 |
|---|---|---|---|---|---|
| $N_S$ | 47 | 18 | 54 | 25 | 41 |
| $N_V$ | 73 | 20 | 54 | 30 | 64 |
| $N_S/N_V$ | 0.64 | 0.90 | 1.00 | 0.86 | 0.64 |

Since MCCNTs are formed by joining two nanotubes of different diameters, the Y-junction is expected to be significantly deformed, as mentioned above. Therefore, the local cross-sectional properties of the MCCNT were investigated. The effective cross-sectional area was obtained numerically using small grids with an area of 0.25 Å$^2$. When an $H_2O$ molecule is adsorbed on the inner surface of a CNT, the O atom and H atoms are 3.3 and 2.2 Å from the nanotube surface, respectively. Thus, we considered the region within 2.2 Angstroms from the CNT surface to be forbidden and exclude the region when calculating the effective cross-sectional area. Fig. 4(a) shows a plot of the effective cross-sectional area of the MCCNT as a function of the *z*-direction (tube axis direction). The red squares in the plot indicate the areas of the local cross-sections estimated from an ellipse. Because the end of the MCCNT is an open part terminated with hydrogen atoms, it bends to form a shape similar to that of a trumpet when it is relaxed. We identified three points (A, B, and C) at which we checked the cross-sectional view, as shown in Fig. 4(b). The shape of the cross-section at the left entrance of the MCCNT slightly deviates from that of an ellipse at Point A. Near

Point B, one can see that the shape obtained by approximating the cross-sectional area is elliptical. However, in the vicinity of the Y-junction (Point C), the shape of the cross-section differs significantly from an elliptical shape.

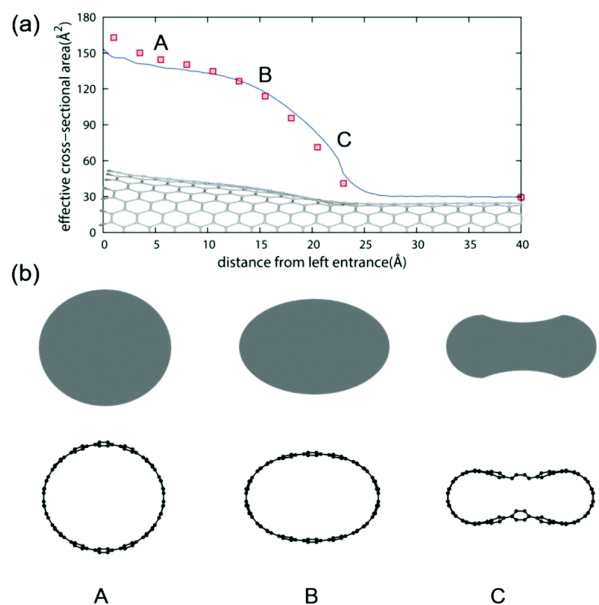

**Fig. 4** (a) Effective cross-sectional area of the MCCNT as a function of the distance from the left entrance. (b) Cross-sectional views for the three selected points (A, B, and C) in (a).

Fig. 5(a) shows the number density of water molecules inside the CNT as a function of the axial distance from the left entrance of the CNT. We found the periodic oscillation of the number density of water molecules with the wavelength of ∼2.5 Å inside the (7, 7) CNT arms along the axial distance. The wavelength is almost the same as the average distance (2.8 Å) between the oxygen atoms in two adjacent water molecules, the so-called hydrogen bond length. The simulation results observed in the present study might show variations depending on the detailed models, including both the CNT structure and water model.[31] Notably, even though there is no noticeable water model dependence within current simulations, there could be none-negligible changes due to polarity at the open ends of CNT.[11] In the perfect (14, 14) CNT, this periodicity evens out significantly as the diameter increased. In contrast, in the perfect (7, 7) CNT, the water density shows a pronounced oscillatory pattern as the axial distance changes from the left entrance of the CNT to the centre of the CNT.[38] The relative density difference between (7, 7) and (14, 14) CNT behaviour is attributed to at least two different filling arrangements of water molecules inside the MCCNT, referred to as "single-file mode" water and "layered mode" water.[1] The size and shape of the water nanostructures depend on the local cross-sectional area of the nanochannel that determines the space in which the water molecules will enter. This causes differences in the oscillatory pattern of the water density. To check if there is a difference in density characteristics depending on the water model, we used the TIP3P and TIP4P models to calculate the molecular water density inside the nanotube. Our results show that the water model dependence is minimal, at least between TIP3P and TIP4P models. In other words, the axis-dependent periodic nature of the water density survives in the (14, 14) CNT region of the MCCNT. For the perfect (14, 14) CNT, the oscillation of the water density in MCCNTs is negligible at both 280 and 300 K. For the perfect (7, 7) CNT, we observed that an oscillation with a small amplitude occurs at 300 K, but a noticeably stronger pattern occurs at 280 K due to high ordering at low temperature. As expected from the oscillatory pattern of water density shown in the perfect (14, 14) CNT, the oscillation of the water density in MCCNTs is also negligible near the larger CNT inlet, but begins a sudden and large oscillation at 17.5 Å, dropping significantly at 18.5 Å, and increases sharply again at $z = 20$ Å, similar to the pattern seen in pure (7, 7) CNTs near $z = 25$ Å.

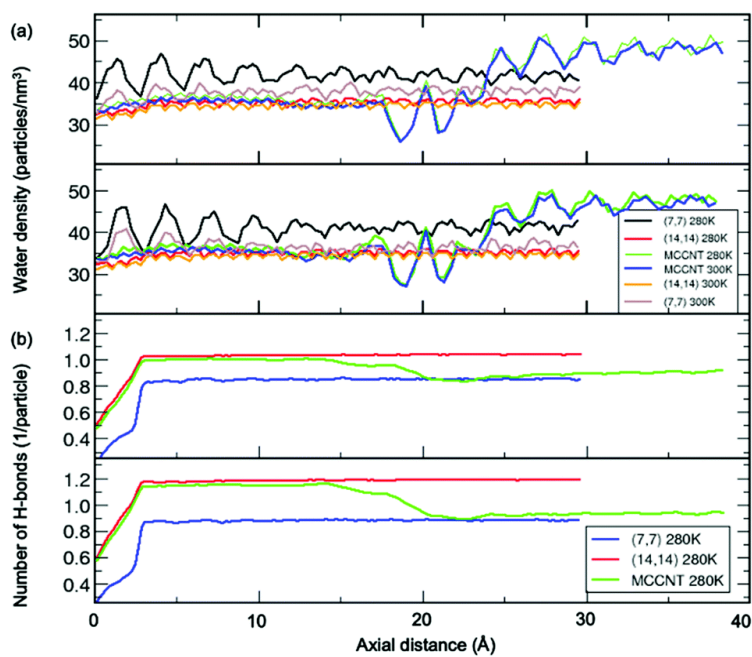

**Fig. 5** (a) Density of water molecules for the MCCNT (14, 14) region as a function of axial distance. The different values in the legend indicate the indices of the CNTs (shown on the right-hand side). (b) Number of hydrogen bonds per water molecule for the perfect (7, 7) and (14, 14) CNTs and the MCCNT as a function of axial distance. Top and bottom panels correspond to the data from TIP3P water and TIP4P water models, respectively.

We conducted a more in-depth investigation of the water molecules inside our MCCNT model, especially in the vicinity of the Y-junction region, by calculating the average number of hydrogen bonds per water molecule as a function of the axial distance. Here, we defined that a hydrogen bond was "formed" when the distance between the O atom of the water molecule and the H atom of the neighbouring water molecule was shorter than 3.5 Å, and the bonding angle ∠O···H–O was within 30°,[40] where ··· and – represent hydrogen and covalent bonds, respectively. In Fig. 5(b), we compare the above results of the

MCCNT with those of the perfect (7, 7) and (14, 14) CNTs at 280 K. The average number of H-bonds at 300 K was smaller than at 280 K, but the overall trend was almost the same. As expected, the extent of H-bonding of water in the (14, 14) CNT part of the MCCNT, resembles that of the perfect (14, 14) CNT. The small number of hydrogen bonds up to $z = 3$ Å appears because we did not consider water molecules outside the CNTs in the calculations. Thus, we will ignore the results near the CNT inlet. In general, the average number of H-bonds increase as the CNT diameter increases.[39,41] The number of H-bonds in the (14, 14) CNT region (the (7, 7) CNT arm) of MCCNTs approaches that in the perfect (14, 14) CNTs (the perfect (7, 7) CNT). Interestingly, the average number of hydrogen bonds in the water molecules inside the MCCNT decreases stepwise from left to right. The average number of H-bonds begins to decrease at $z = 17.5$ Å, far before reaching the Y-junction.

Here arise three interesting questions on the water molecules confined inside the MCCNT: (1) Why does the water density begin to show high amplitude oscillation from around $z = 17.5$ Å? (2) Why does the number of H-bond begin to decrease from around the same region? (3) Why does the average number of hydrogen bonds in the (7, 7) CNT region (Region 3 in Fig. 3) slowly increases in MCCNT? Now, we answer the three questions above. The left Y-junction is near $z = 24$ Å. We emphasize severe structural deformation near the Y-junction (Region 2) where the CNTs with different diameters are interconnected. Very interestingly, in Regions 2 and 3, water molecules are linked together by hydrogen bonds in a closed-loop, as clearly shown in Fig. 6. By the local deformation of the water nanostructure, a unique H-bonded network forms in Region 2 where the elliptical column-shaped water nanostructure is connected to the flat, double-stranded water in Region 3, as clearly shown in the side view in Fig. 3. Such a unique H-bonds in Region 2 makes the large-amplitude oscillation in the water density begin at $z = 17.5$ Å (at the middle of Region 2). For the same reason, the number of H-bonds also changes at this position. The deformed shape of the tube in Region 2 affects the arrangement of the water molecules in the host tube, which does not allow the double-stranded water to twist within the (7, 7) CNT arm in Region 3 unlike the perfect (7, 7) CNT, where a water strand can form a helix. The cross-sectional view of the MCCNT along the line M2 in Fig. 1 shows slight distortion from the perfect (7, 7) CNT circular form, which should contribute to the ordering of water molecules in it. In other words, in combination with the slight structural distortion in Region 3, the water molecules in Region 2 that are connected to a water network within Region 3 make the most of $H_2O$ molecules to be placed on a planar form in the (7, 7) CNT arms (Region 3). As a result, the water molecule network inside the (7, 7) CNT arms of the MCCNT has higher hydrogen bond ordering, and the average number of H-bonds increases gradually, compared with the case of the perfect (7, 7) CNT.

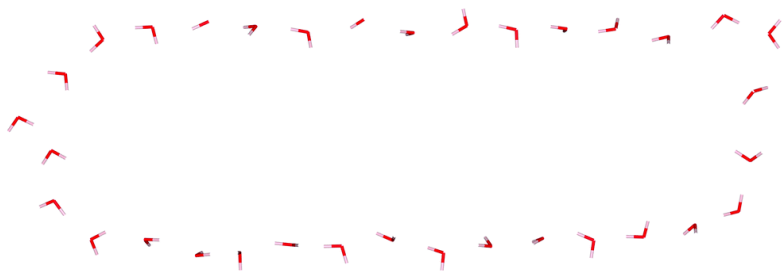

**Fig. 6** The MCCNT simulation snapshot at 280 K with TIP3P water, showing the formation of a water network in the (7, 7) CNT arm with waters in Region 2.

Finally, we need to mention that the results could be somewhat different depending on the model force field even though water model dependence is minimal here, within TIP3P and TIP4P water. But it is noted that the density of water along the tube axis and the average number of hydrogen bonds inside (7,7) CNT obtained from current model for is almost identical to the results (about 0.8 per water molecule)[39] from simulations using Brenner potential[42] based on Tersoff's covalent-bonding formalism, suggesting that general feature of current simulation, which is the structure of water molecules within CNT, might be largely valid.

**Conclusions**

In summary, we have studied hydrogen-bonded nanostructures of water formed inside a multiply connected carbon nanotube in dynamic equilibrium in a water bath, using molecular dynamics simulations, and calculated the density of water molecules and the number of hydrogen bonds per $H_2O$ molecule. We found large oscillatory patterns in the density and the average number of hydrogen bonds near the Y-junction of the MCCNT. Remarkable changes occur in the (14, 14) CNT region approximately 6–7 Å before the Y-junction in terms of the density and the ordering of the water molecules. The increased number of hydrogen bonds in the (7, 7) CNT arm of the MCCNT would be expected to eventually approach that of pure (7, 7) CNT as the distance increases. In brief, the present study shows the structural features of water molecules confined inside an inhomogeneous CNT, an MCCNT to be more specific, does not necessarily reflect its immediate confining environment. Instead, as a result of rather complicated interactions, the local cross-sectional area of the host tube can affect water nanostructure over a quite long distance, say up to 1 nm.

Understanding both the dynamic and static behaviours of water inside a topologically complex nanostructure is of fundamental importance because of their numerous possible applications in biology, catalyst engineering, and nanofluidics.[43] Recent experiment shows the population increase of the water network confined in the mesoporous silica.[43] In this work, we presented the static properties of water within the symmetric MCCNT consisting of two Y-junctions. Although the carbon nanotubes with a branched structure attract much interest mainly because of its possible applications as nano-electronic devices due to unique electronic properties,[44] it has been pointed out that the structure also has a high potential to be used as a molecular sieve or molecular absorbent, including desalination of water, by reducing the pressure to drive the corresponding process.[45] In this sense, the detailed behavior of water molecules within the CNT with Y-junctions has fundamental interests. We believe that our work will give better insight and information on how the water fluid behaves in the structurally complex nanochannels.

## Acknowledgements

The authors were supported by the Mid-career Researcher Program (Grant No. NRF-2016R1A2B2016120) through the National Research Foundation funded by the Ministry of Science and ICT of Korea. SJ was in part supported Basic Research Program (Grant No. 2012R1A1A2009242 & 2017M3D9A1073784) by the National Research Foundation of Korea.

## Reference

[1] A. Alexiadis and S. Kassinos, Chem. Rev., 2008, 108, 5014–5034

[2] G. Hummer, J. C. Rasaiah and J. P. Noworyta, Nature, 2001, 414, 188 CrossRef CAS PubMed .

[3] T. A. Pascal, W. A. Goddard and Y. Jung, Proc. Natl. Acad. Sci. U. S. A., 2011, 108, 11794–11798

[4] S. Chakraborty, H. Kumar, C. Dasgupta and P. K. Maiti, Acc. Chem. Res., 2017, 50, 2139–2146

[5] S. Dalla Bernardina, E. Paineau, J.-B. Brubach, P. Judeinstein, S. p. Rouzière, P. Launois and P. Roy, J. Am. Chem. Soc., 2016, 138, 10437–10443

[6] M. Gordillo and J. Martı, Chem. Phys. Lett., 2000, 329, 341–345

[7] I. Hanasaki and A. Nakatani, J. Chem. Phys., 2006, 124, 174714

[8] M. F. Chaplin, in Adsorption and phase behaviour in nanochannels and nanotubes, Springer, 2010, pp. 241–255

[9] R. Das, M. E. Ali, S. B. A. Hamid, S. Ramakrishna and Z. Z. Chowdhury, Desalination, 2014, 336, 97–109

[10] S. Joseph and N. Aluru, Nano Lett., 2008, 8, 452–458

[11] S. K. Kannam, B. Todd, J. S. Hansen and P. J. Daivis, J. Chem. Phys., 2013, 138, 094701

[12] M. H. Köhler, J. R. Bordin, C. F. de Matos and M. C. Barbosa, Chem. Eng. Sci., 2019, 203, 54–67

[13] S. K. Kannam, P. J. Daivis and B. Todd, MRS Bull., 2017, 42, 283–288

[14] A. J. McGaughey and D. Mattia, MRS Bull., 2017, 42, 273–277

[15] D. Erickson, D. Li and C. Park, J. Colloid Interface Sci., 2002, 250, 422–430

[16] H. Mehrabian, P. Gao and J. J. Feng, Phys. Fluids, 2011, 23, 122108

[17] D. Shou, L. Ye, J. Fan and K. Fu, Langmuir, 2013, 30, 149–155

[18] D. Shou, L. Ye, J. Fan, K. Fu, M. Mei, H. Wang and Q. Chen, Langmuir, 2014, 30, 5448–5454

[19] B. Figliuzzi and C. Buie, J. Fluid Mech., 2013, 731, 142–161

[20] J. Berthier, D. Gosselin, A. Pham, G. Delapierre, N. Belgacem and D. Chaussy, Langmuir, 2016, 32, 915–921

[21] S. Bekou and D. Mattia, Curr. Opin. Colloid Interface Sci., 2011, 16, 259–265

[22] K. Ritos, D. Mattia, F. Calabrò and J. M. Reese, J. Chem. Phys., 2014, 140, 014702

[23] G. Kim, S. B. Lee, T.-S. Kim and J. Ihm, Phys. Rev. B: Condens. Matter Mater. Phys., 2005, 71, 205415

[24] M. Yoon, S. Han, G. Kim, S. B. Lee, S. Berber, E. Osawa, J. Ihm, M. Terrones, F. Banhart and J.-C. Charlier, Phys. Rev. Lett., 2004, 92, 075504

[25] B. Satishkumar, P. J. Thomas, A. Govindaraj and C. Rao, Appl. Phys. Lett., 2000, 77, 2530–2532

[26] J. Romo-Herrera, M. Terrones, H. Terrones, S. Dag and V. Meunier, Nano Lett., 2007, 7, 570–576

[27] L. P. Rajukumar, M. Belmonte, J. E. Slimak, A. L. Elías, E. Cruz-Silva, N. Perea-López, A. Morelos-Gómez, H. Terrones, M. Endo and P. Miranzo, Adv. Funct. Mater., 2015, 25, 4922

[28] P. J. Harris, I. Suarez-Martinez and N. A. Marks, Nanoscale, 2016, 8, 18849–18854

[29] D. Van Der Spoel, E. Lindahl, B. Hess, G. Groenhof, A. E. Mark and H. J. Berendsen, J. Comput. Chem., 2005, 26, 1701–1718

[30] W. L. Jorgensen, J. Chandrasekhar, J. D. Madura, R. W. Impey and M. L. Klein, J. Chem. Phys., 1983, 79, 926–935

[31] W. L. Jorgensen, D. S. Maxwell and J. Tirado-Rives, J. Am. Chem. Soc., 1996, 118, 11225–11236

[32] Y. Nakamura and T. Ohno, Mater. Chem. Phys., 2012, 132, 682–687

[33] G. Hummer, J. C. Rasaiah and J. P. Noworyta, Nature, 2001, 414, 188

[34] I. Hamada, Phys. Rev. B: Condens. Matter Mater. Phys., 2012, 86, 195436

[35] T. Kurita, S. Okada and A. Oshiyama, Phys. Rev. B: Condens. Matter Mater. Phys., 2007, 75, 205424

[36] O. Leenaerts, B. Partoens and F. Peeters, Phys. Rev. B: Condens. Matter Mater. Phys., 2009, 79, 235440

[37] J. Ma, A. Michaelides, D. Alfe, L. Schimka, G. Kresse and E. Wang, Phys. Rev. B: Condens. Matter Mater. Phys., 2011, 84, 033402

[38] Y. Yang, F. Liu and Y. Kawazoe, J. Phys. Chem. Solids, 2019, 124, 54–59

[39] Z. Xiao-Yan and L. Hang-Jun, Chin. Phys., 2007, 16, 335

[40] A. Soper and M. Phillips, Chem. Phys., 1986, 107, 47–60

[41] R. J. Mashl, S. Joseph, N. R. Aluru and E. Jakobsson, Nano Lett., 2003, 3, 589–592

[42] D. W. Brenner, Phys. Rev. B: Condens. Matter Mater. Phys., 1992, 46, 1948

[43] A. W. Knight, N. G. Kalugin, E. Coker and A. G. Ilgen, Sci. Rep., 2019, 9, 8246

[44] M. Zhang and J. Li, Mater. Today, 2009, 12, 12

[45] Y. Zhang, Int. J. Heat Mass Transfer, 2017, 114, 536